\documentclass[prb,superscriptaddress, preprint,showpacs,showkeys,amsmath]{revtex4}

\usepackage{amssymb}
\usepackage{amsmath}
\usepackage{graphicx}
\begin{document}


\title{Anomalous low-temperature magnetoelastic properties of nanogranular (CoFeB)$_{x}$-(SiO$_{2}$)$_{1-x}$}


\author{A.A. Timopheev}
\email[e-mail: ]{ timopheev@iop.kiev.ua}
\affiliation{Institute of Physics NAS of Ukraine, Prospect Nauki str. 46, Kiev, 03028, Ukraine}

\author{S.M. Ryabchenko}
\affiliation{Institute of Physics NAS of Ukraine, Prospect Nauki str. 46, Kiev, 03028, Ukraine}

\author{V.M. Kalita}
\affiliation{Institute of Physics NAS of Ukraine, Prospect Nauki str. 46, Kiev, 03028, Ukraine}

\author{A.F. Lozenko}
\affiliation{Institute of Physics NAS of Ukraine, Prospect Nauki str. 46, Kiev, 03028, Ukraine}

\author{P.A. Trotsenko}
\affiliation{Institute of Physics NAS of Ukraine, Prospect Nauki str. 46, Kiev, 03028, Ukraine}

\author{V.A. Stephanovich}
\affiliation{Institute of Mathematics and Informatics, Opole University, Oleska 48, 45-052 Opole, Poland}

\author{A.M. Grishin}
\affiliation{Royal Institute of Technology, Electrum 229, S-164 40 Kista, Stockholm, Sweden}

\author{M. Munakata}
\affiliation{Energy Electronics Laboratory, Sojo University, Kumamoto 860-0082, Japan}


\date{\today}

\begin{abstract}
We report magnetostatic measurements for granulated films (CoFeB)$_{x}$-(SiO$_{2}$)$_{1-x}$ with fabrication induced intraplanar anisotropy. The measurements have been performed in the film plane in the wide temperature interval 4.5$\div$300 K. They demonstrate that above films have low-temperature anomaly below the percolation threshold for conductivity. The essence of the above peculiarity is that below 100 K the temperature dependence of coercive field for magnetization along easy direction deviates strongly from Neel-Brown law.  At temperature lowering, the sharp increase of coercivity is observed, accompanied by the appearance of coercive field for magnetization along hard direction in the film plane. We establish that observed effect is related to the properties of individual ferromagnetic granules. The effect weakens as granules merge into conglomerates at $x$ higher then percolation threshold and disappears completely at $x>1$. We explain the above effect as a consequence of the difference in thermal expansion coefficients of granule and cover material. At temperature lowering this difference weakens the envelopment of an individual granule by the cover matrix material, thus permitting to realize the spontaneous magnetostriction of a granule. The latter induces an additional anisotropy with new easy axis of a granule magnetization along the external magnetic field direction. Our explanation is tested and corroborated by the ferromagnetic resonance measurements in the films at $T$ = 300 K and $T$ = 77 K.
\end{abstract}


\maketitle

\section{\label{sec:intro}Introduction}
The nanogranulated ferromagnets are composite materials, consisting of ferromagnetic granules embedded in nonmagnetic matrix. The average granules size in such composites (below conductivity percolation threshold) lies in nanometric range. The granulated structures have high magnetization and magnetization reversal speed. The matrix material can be either non-magnetic metal or dielectric. In the latter case, the structures have high resistivity and demonstrate magnetoresistance which makes these structures to be promising as magnetic sensors. The typical representatives of this class of materials are granulated ferromagnetic films (CoFeB)$_{x}$-(SiO$_{2}$)$_{1-x}$ \cite{r1,r2,r3,r4}. In such systems, giant magnetoresistance is observed at $x$ below percolation threshold \cite{r5}, the anisotropic magnetoresistance is observed above this threshold \cite{r6} and their joint manifestation occurs close to the threshold \cite{r7}, making these films to be promising as magnetoresistive sensors.

The majority of studies of above composites are related to their high-frequency and magnetotransport properties at room temperature as their possible applications are expected in latter temperature range. Nevertheless, their low-temperature magnetostatic investigations can be also interesting, although they are poorly represented in a literature. The number of low-temperature studies has been devoted to the effect of intergranular magnetic interactions and corresponding correlation of granules magnetic moments directions below certain temperature. The great deal of available low-temperature studies of nanogranulated films is devoted to the effects of blocking of thermoactivated magnetization reorientation in monodomain ferromagnetic granules. Latter effects generate the temperature dependence of films coercive field and remnant magnetization, which is a bit different for different types of magnetic anisotropy. The discussion of temperature dependence of films coercive field and remnant magnetization, unrelated to the above blocking is almost absent in a literature. In particular, to the best of our knowledge, there are no papers discussing the low-temperature manifestations of magnetoelastic interactions in magnetostatic and/or magnetoresonant properties of granulated ferromagnetic films. Actually, there is quite small number of papers devoted to the magnetoelastic properties of the above films. We think that this is related to the fact, that according to Ref.~\onlinecite{r8}, the magnetoelastic effects in a composite are much weaker then those in a bulk ferromagnet. Below we show that in spite of the above, the magnetoelastic effects can generate nontrivial low-temperature anomalies in the physical properties of nanogranulated ferromagnetic films.

In the granular films ((CoFeB)$_{x}$-(SiO$_{2}$)$_{1-x}$, which will be considered below, the uniaxial intraplanar anisotropy in the film plane appears in the process of fabrication. This anisotropy is the same for each granule and conserves below percolation threshold. Also, the fabrication technology permits to obtain very small dispersion of granule sizes. This means that below percolation threshold such granular system can be considered as an easy-axis oriented ensemble of Stoner-Wolfarth (SW) particles \cite{r9}. According to Neel-Brown law \cite{r10}, for such ensemble, the coercive field depends on temperature as a square root for magnetization along the easy axis:

\begin{equation}\label{EQ1}
 H_{c}(T)=H_{c}(0)(1-\sqrt{T/T_{b}^{NB}}),
\end{equation}
where

\begin{equation} \label{EQ2}
 T_{b}^{NB} =\frac{K\cdot V}{k\ln (t\cdot f_{0})},\text{ and } H_{c}(0)=\frac{2K}{m_{gr}}=H_{a}.
\end{equation}

Here $T$ is a temperature, $K$ is the anisotropy constant, $V$ is a particle volume, $m_{gr}$ is its magnetization, $k$ is Boltzmann constant, $t$ is a measuring time,  $f_{0}$ is a factor characterizing the attempt frequency in Arrhenius law (the usual range here is $10^{8}\div10^{12} s^{-1}$), $T_{b}^{NB}$  is so-called blocking temperature. Latter quantity is a boundary between temperature range $T<T_{b}^{NB}$, where the state of the granules ensemble is metastable (blocked) during measurement time $t$ with respect to thermally activated reorientations of granules magnetic moments and $T>T_{b}^{NB}$, where the system is in equilibrium (unblocked) state. For a magnetization of oriented SW particles ensemble along hard direction the coercivity should be absent. Note that Eq.\ref{EQ1} is derived in the framework of hypothetical relaxational model and does not describe precisely the concept of measuring time used in magnetostatic experiments. Also, this equation is obtained in low-temperature approach. It does not work well in the temperature range close to $T_{b}^{NB}$, see Refs.~\onlinecite{r11,r12} for details. However, in spite of these drawbacks, in the majority of cases it is suitable for determination of the anisotropy field $H_{a}$ and blocking temperatures from the dependencies $H_{c}(T)$.

Several remarks are in place here. Namely, the temperature dependence of granules saturation magnetization can modify dependence \eqref{EQ1}. One of the ways to consider such modification has been suggested in Ref.~\onlinecite{r13}. The modification of dependence \eqref{EQ1} can also be related to the temperature dependence of the magnetic anisotropy constant of granules material. The other remark is that the approach of SW particles ensemble is invalid for the case of sufficiently strong interparticle magnetic interaction. In this case, the character of magnetization curves can alter substantially. Namely, the coercivity, related to "superferromagnetic" \cite{r14} or "superspinglass" \cite{r15} states can appear.

 In the present paper, we report the results of magnetostatic measurements for series of nanogranulated films $(CoFeB)_{x}-(SiO_{2})_{1-x}$ with different $x$ in the wide temperature range from those close to $T_{b}^{NB}$ till substantially lower ones (from 300 K till 4.5 K). The measurements are performed for magnetization in the film plane along both easy and hard (in the plane) magnetization directions. We will show that at $T<100$K the dependence $H_{c}(T)$ for magnetization along easy direction deviates strongly from Neel-Brown law. At temperature lowering, the sharp increase of coercivity along easy direction is observed, accompanied by the appearance of coercive field for magnetization along hard direction in the film plane. The similar to this low-temperature behavior has not been described in the literature earlier. We explain this behavior as a manifestation of magnetoelasticity of ferromagnetic granules under change of their envelopment by matrix material at temperature variations. In other words, we explain such behavior as a consequence of granulated structure of such films. This hypothesis will be checked by ferromagnetic resonance (FMR) measurements in these films.

 \section{\label{sec:Experimental}Experimental details}

The granulated ferromagnetic films $(Co_{0.25}Fe_{0.66}B_{0.09})_{x}-(SiO_{2})_{1-x}$  on glassy substrate were grown in the University of Sojo (Japan) by the method of magnetron sputtering of magnetic and nonmagnetic components on substrate, secured on the rotated and cooled drum. As a result, the interplanar uniaxial anisotropy has been induced in the film plane. The films thickness is about 500 nm. The details of above technology can be found in Ref.~\onlinecite{r16}. The results of x-ray studies \cite{r7} show that the fabricated films have amorphous ferromagnetic granules with almost spherical shape. The magnetostatic measurements have been carried out on the vibrating sample magnetometer LDJ-9500. The cryostat with temperature stabilisation on the base of gaseous He blow into the working cavity have been used for investigations in the temperature interval $4.5<T<295$ K. The accuracy of temperature stabilization was $\pm$1 Ê. The FMR measurements at $T=295$ and 77 K have been carried out on the X-band radiospectrometers RADIOPAN SE/X-2544 and Bruker - E 500 CW. The FMR measurements under controlled uniaxial film compression along with substrate in its plane have been carried out at room temperature ($T=295$ K) only. The compression has been carried out by a special device for pressure creation \cite{r17}. The films with $x$ both above and below percolation threshold have been studied. The film with $x\rightarrow1$ has also been investigated, both on the substrate and without it.

 \section{\label{sec:Magnetostatic}The results of magnetostatic measurements}

\begin{figure}
\begin{center}
\includegraphics*[width=8cm]{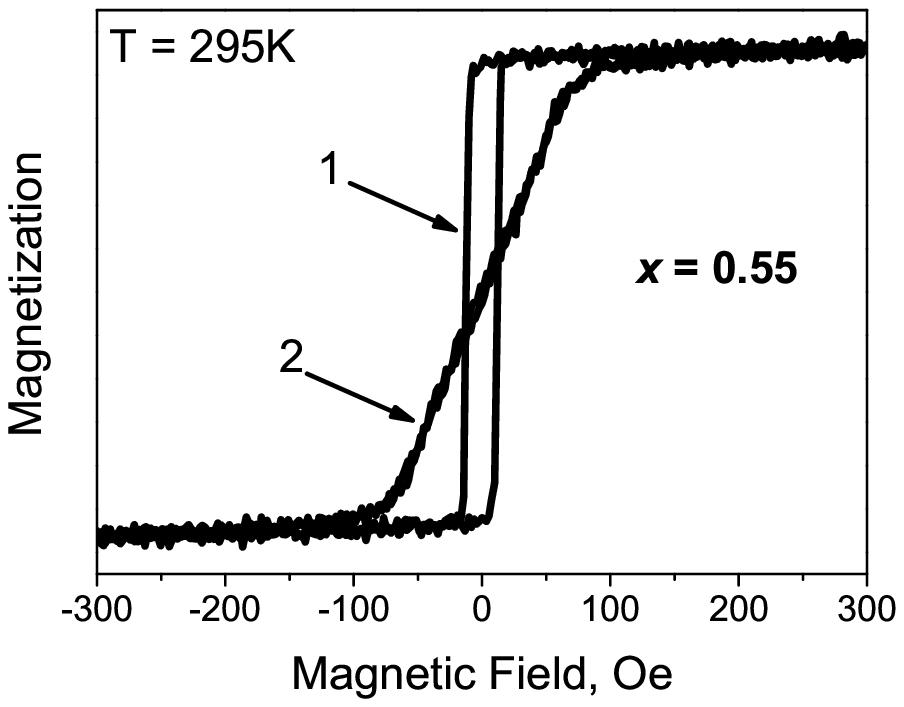}
\end{center}
\caption{\label{fig1} The magnetization curves for the sample with $x=0.55$ at $Ò=300$ K. Curve 1 corresponds to easy direction , curve 2 to hard direction in the film plane.}
\end{figure}

The results of magnetostatic measurements for the sample with $x=0.55$ along easy (curve 1) and hard (curve 2) directions in the film plane at $T=295$ K are reported on Fig. \ref{fig1}. It is seen, that there is a hysteresis along easy direction with almost $100\%$ remanent magnetization, while there is no hysteresis with almost linear magnetic field dependence of magnetization along hard direction. Latter curve has a cusp at magnetic field $H=H_{a}=87\pm5$ Oe. At $-H_{a}>H$ and $H>H_{a}$ the magnetization curve saturates. This behavior is typical for ensemble of easy-axis oriented monodomain granules. Thus, the film can be considered as an ensemble of oriented SW particles. From the measurements at different temperatures, we have determined that the saturation magnetization value does not change in the wide temperature interval $4.5<T<300$ K.  The effect of giant magnetoresistance has also been observed in this film. The magnetoresistance is independent of mutual orientation of magnetic field $H$ and measuring current. This means that this film consists of metallic granules, isolated from each other by dielectric layer, i.e. granules concentration $x$ in the film lies below percolation threshold.

As it has already been mentioned in the Section \ref{sec:intro}, the dependence $H_{c}(T)$ for the above system at $T<T_{b}^{NB}$ should have square root character for the magnetization along easy axis. Its extrapolation to $T\rightarrow0$ (point of intersection with ordinate axis) determines $H_{a}$, while the extrapolation to $H_{c}\rightarrow0$ (point of intersection with abscissa axis) - $T_{b}^{NB}$. The anisotropy field $H_{a}$, obtained by the above extrapolation procedure, should correspond to that obtained from magnetostatic measurements along hard direction in the plane.

\begin{figure}
\begin{center}
\includegraphics*[width=16cm]{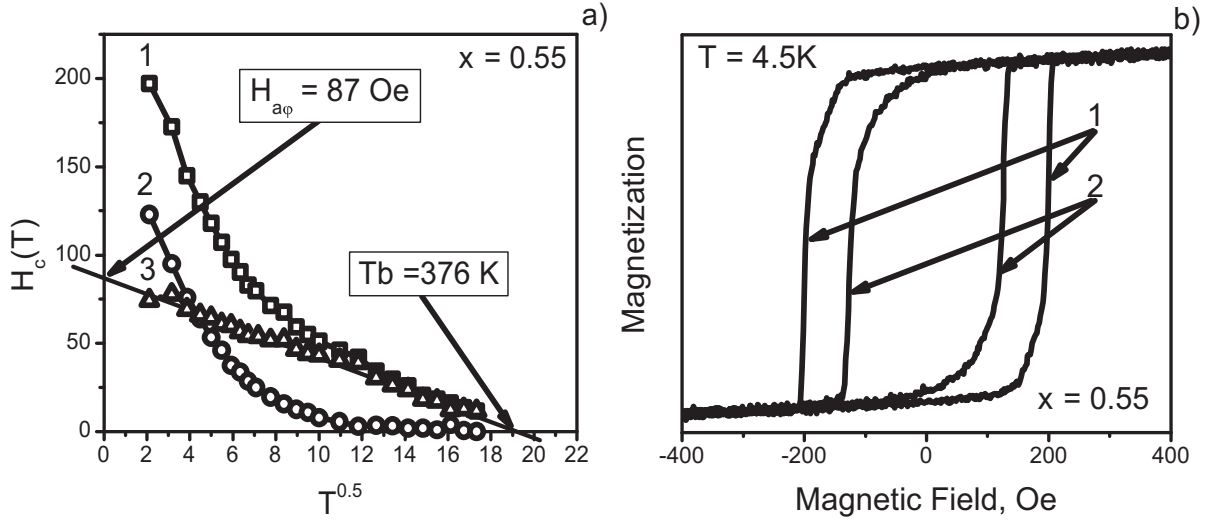}
\end{center}
\caption{\label{fig2} Panel a)-- temperature dependences of coercive field for the sample with $x=0.55$. Curve 1 corresponds to the measurements along easy axis; curve 2 -- along hard axis in the film plane, curve 3 is a difference between curves 1 and 2. Full straight line gives an extrapolation of curve 3. Panel b) -- magnetization curves at $Ò=4.5$ Ê along easy (curve 1) and hard (curve 2) directions in the film plane.}
\end{figure}

The dependence $H_{c}(T)$ for measurements along easy direction in the film plane is depicted in the square root temperature scale on Fig. \ref{fig2}a (curve 1). It is seen that this dependence can be described by Neel-Brown law in the temperature range $100\div295$ K. In this interval, the above extrapolation procedure determines the anisotropy field $H_{a}=90\pm5$ Oe, close to that obtained from measurements along hard direction at $T=295$ K (see Fig. \ref{fig1}). The blocking temperature $T_{b}$, obtained from the extrapolation procedure for above sample, turns out to be $T_{b}=T_{b}^{NB}=376\pm5$ K. The magnetostatic measurements \cite{r18}, fulfilled up to higher temperatures ($\approx500$ K) corroborated the latter $T_{b}$ value. At temperature lowering below 100 K the coercive field growth becomes more intensive so that dependence $H_{c}(T)$ deviates from Eq. \eqref{EQ1} with above values $T_{b}$ and $H_{c}(0)=H_{a}$, giving coercive field $Hc\approx200$ Oe at $Ò=4.5$ Ê (Fig. \ref{fig2}b, curve 1).

More surprisingly, for the measurements along hard (in the film plane) direction at $Ò<100$ K the magnetization curves modify drastically. They are no more typical for an ensemble of oriented SW particles, magnetized along hard direction. Namely, instead of linear anhysteretic magnetization curve at $Ò>100$ K (see curve 2 on Fig. \ref{fig1}), one can observe hard hysteretic curve with almost 100\% remanence (Fig. \ref{fig2}b, curve 2 for $T=4.5$ K). We note here, that Neel-Brown model does not imply the above modification of magnetization curves with temperature lowering.

The curve 3 on Fig. \ref{fig2}a reports the difference of curves 1 and 2 (from the same figure), showing $H_{c}(T)$ for magnetization along easy and hard directions respectively. It turns out, that just the difference of curves 1 and 2 satisfies the Neel-Brown law down to $T\rightarrow0$ K. Moreover, its extrapolation to $Ò=0$ gives $H_{a}$ which is close to that from curve 2 of Fig. \ref{fig1}, obtained at $T=295>100$ K or from extrapolation to $T=0$ K of linear part of curve 1 on Fig. \ref{fig2}a, obtained at $T>100$ K.

\begin{figure}
\begin{center}
\includegraphics*[width=8cm]{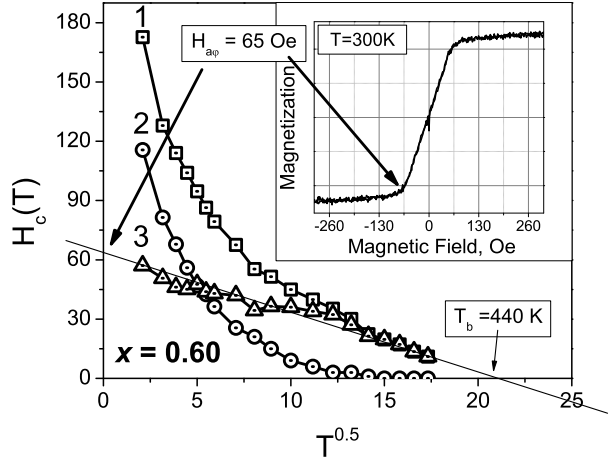}
\end{center}
\caption{\label{fig3} Temperature dependences of coercive field for the sample with $x=0.60$. Curves 1-3 and full straight line are similar to those in Fig.\ref{fig2}. Inset shows the magnetization curve for hard (in the film plane) direction magnetization at $T=295$ K.}
\end{figure}

The observed effect is not peculiar to the only one film with $x=0.55$. The same measurements for the sample with little larger ferromagnetic component content ($x=0.60$) give the similar result (see Fig. \ref{fig3}). For latter sample, the blocking temperature is somewhat higher ($T_{b}=440$ K), that can be explained by a bit larger granules size.  The anisotropy field $H_{a}$ for latter sample is a little smaller then that for the sample with $x=0.55$. The rest of features of magnetization curves for the sample with $x=0.60$ have the same tendency as those for the sample with $x=0.55$.

The peculiarity of the sample with $x=0.70$ is that its magnetic component content is higher then percolation threshold so that the part of granules form the continuous conducting channels. That is why in this sample, contrary to two previous samples, the mixture of giant and anisotropic magnetoresistances is observed. The latter fact means that the part of granules is still electrically isolated and these granules take the same part in conductivity as those in continuous conducting channels. It is reasonably to suppose that in this sample both ``conductivity percolation'' and ``magnetic percolation'' take place. Latter kind of percolation means that the directions of magnetic moments of neighboring granules are no more independent, but they are rather correlated due to intergranular spin-spin interaction or dipole scattering fields. The dependencies $H_{c}(T)$ for easy and hard (in the film plane) directions do not have features observed in the samples with $x=0.55$ and $x=0.60$. This is clear since the percolated (primarily "magnetically percolated") sample cannot be described by Neel - Brown theory as an ensemble of noninteracting SW particles.

\begin{figure}
\begin{center}
\includegraphics*[width=8cm]{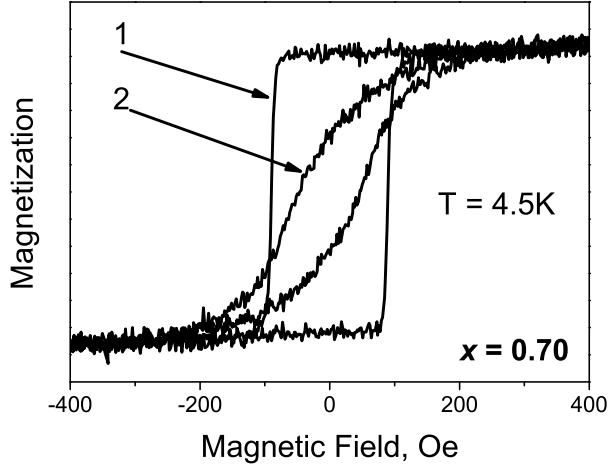}
\end{center}
\caption{\label{fig4} The magnetization curves for the sample with $x=0.70$ at $T=4.5$ K. Curve 1 corresponds to easy direction, curve 2 to hard direction in the film plane.}
\end{figure}

Despite that, at low temperatures, the coercive field along hard (in the film plane) direction appears also (see curve 2 on Fig. \ref{fig4}), although its value is much smaller then that for the films with $x=0.55$ and $x=0.60$. The value of remanent magnetization for this direction at $T=4.5$ K, which equals to around 40\% of saturation magnetization, permits us to conclude that the same percentage of ferromagnetic component in the sample is left as isolated granules. These granules are indeed responsible, similar to above cases of $x=0.55$ and $x=0.60$, for discussed low-temperature anomaly.   The confirmation of this conclusion is the fact that in the samples with even higher (higher then $x=0.70$) concentrations (actually up to $x\rightarrow1$), there is no coercivity along hard (in a plane) direction and no anomalous growth of coercivity along easy direction up to $T=4.5$ K. The entire magnetization curve for such samples does not alter at lowering temperature.

Thus, the fulfilled complex of magnetostatic measurements permits to conclude that above anomalous low-temperature effect in coercivity is observed in the samples below and close to percolation threshold, i.e. it is related to the properties of isolated granules.

\section{\label{sec:Disscussion}Discussion and possible model}

The first suggestion about possible physical reason of observed low-temperature anomaly was that the properties of magnetic material alter with temperature lowering. However, as it has already been mentioned above, this effect is substantially weakened in the films with $x$ higher then percolation threshold and completely disappears in a film with $x\rightarrow1$. We consider this concentrational behavior to be the main argument in the discussion of possible physical reasons of the above low-temperature anomaly. Namely, it makes us to conclude, that this effect is not a property of granules ferromagnetic material. One can also suppose that the partial ``decoherence'' of granules easy axes (due to certain technological details) in a plane occurs at low temperatures. This might lead to the appearance of coercivity along hard direction. But such ``decoherence'' yield the lowering of remanent magnetization. However, it is seen from Fig. \ref{fig2}b that hysteresis loops have 100\% remanence for both above directions at $T=4.5$ K, which is possible only in the case when a magnetic field is parallel to the easy axes of all (or almost all) granules.

Since Neel-Brown law is still valid for difference curve $H_{c}(T)$, we might suppose that at temperature lowering an additional anisotropy appears. This hypothetical anisotropy should be induced both along easy direction, giving rise to deviations from Neel-Brown law and along hard direction, yielding the appearance of coercivity. In this case, the initial technological anisotropy (with easy and hard directions in a film plane), responsible for the shape of magnetization curves for nonpercolated films at $T>100$ K, is conserved, yielding the observed temperature dependence of difference curve, following Neel-Brown law. To check this, we have carried out the calculations of hysteresis loops by recursive method \cite{r11} for various cases of joint action of different contributions into granules magnetic anisotropy. First contribution was uniaxial and proportional to $K\cdot sin^{2}(\theta- \theta_{0})$ with fixed direction of $\theta_{0}$ axis and old constant $K$ value (``old'' anisotropy). Second contribution had temperature dependent anisotropy constant and random axes orientation. Namely, we have considered ``randomly oriented'' uniaxial anisotropy of $K_{rand}\cdot sin^{2}(\theta-\theta_{rand})$ type, uniaxial anisotropy of $K_{rand}\cdot sin^{4}(\theta-\theta_{rand})$ type and cubic anisotropy both with fixed and randomly oriented directions of its main axes. These calculations have not given the behavior of hysteretic curves, even qualitatively corresponding to those observed experimentally. To achieve such correspondence we need that easy axis of the additional anisotropy should always (at any magnetization direction in a film plane) be oriented along magnetic field direction.

One of the possible ways of qualitative explanation of observed anomaly is a hypothesis about spontaneous granules magnetostriction generated in the magnetization direction at temperature lowering. The manifestations of such magnetostriction for magnets with spatially degenerate magnetic moment directions (like easy-plane ferro- and antiferromagnets) have been considered in Refs.~\onlinecite{r19,r20,r21}. It has been shown there, that the above magnetostriction generates anisotropy with easy axis, which is always oriented along magnetization direction and rotated at magnetization rotations.

In our case, we should note that such possibility is absent for granules at high temperatures since each of them is tightly enveloped by matrix material. It is possible that there exist only some part of "initial" spontaneous magnetostriction as granules easy axes are technologically oriented. This effect reveals as small intraplanar anisotropy. However, the thermal expansion coefficient of matrix material ($SiO_{2}$) is lower then that of metallic ferromagnetic granule ($CoFeB$).  This difference should yield the partial or complete release of granules from glassy matrix envelopment at temperature lowering and to the possibility of magnetostriction. At granules concentration growth above percolation threshold, they will merge with each other so that for such large clusters the magnetostriction possibilities are limited by its mechanical coupling to substrate. The more lateral dimension has such cluster, the less magnetostriction possibilities it has. At $x\rightarrow1$ this mechanical coupling inhibits completely the possibility of intraplanar magnetostriction for such film.

At high temperatures, when granules are clumped into glassy matrix, the pressure of this matrix compensates the pressure due to magnetostriction so that the granules remain without magnetostriction. At temperature lowering the granules release from the matrix clamping (envelopment) so that the spontaneous magnetostriction can not only appear, but reorient. Under magnetization along easy direction the anisotropy will enhance (as compared to ``initial'' case), while under magnetization along hard direction its axis will rotate towards the direction of granule magnetization vector.  Therefore, if below some temperature the number of such ``released granules'' becomes sufficient, then under the magnetization along easy direction the magnetostrictive anisotropy will be added to shape anisotropy. In a hard direction, this effect leads to the compensation of shape anisotropy and to appearance of new easy axis of a granule along the direction of magnetic field. Latter causes the reorientation of magnetic moment and spontaneous magnetostriction axis so that the field direction is always ``easy''.

 For the discussed mechanism to realize, the relation between the time of granule spontaneous magnetostriction reorientation (more precisely the time of elastic equilibrium establishment in a granule at its magnetization reorientation) and the time of its magnetization thermal reorientations is important. The time of elastic equilibrium establishment in a granule with respect to the value of sound velocity in its material ($\sim10^{5} cm/s$) and its linear dimensions ($\sim10^{-6} cm$) can be estimated as $10^{-11} s$. The time of granule magnetization thermal reorientations should be of the same order as $f_{0}^{-1}$ value -- inverse ``attempts frequency'' in Arrhenius law for thermally activated hops in a potential profile of magnetic particle energy (see, e.g. Ref.~\onlinecite{r10}). Usual estimate for $f_{0}^{-1}$ is $10^{-8}\div10^{-12} s$. This estimation shows that the spontaneous magnetostriction reorientation of a particle can be faster then its magnetization thermal reorientation. In this case the anisotropy related to spontaneous magnetostriction will not stabilize (with respect to thermal fluctuations) the direction of a granule magnetic moment so that our mechanism should not take place\cite{r22}. One can suppose, however, that granules do not become completely free from matrix clamping at temperature lowering, in which case the time of granule spontaneous magnetostriction reorientation grows substantially and thermal fluctuations of magnetic moment will not alter the spontaneous magnetostriction orientation. In other words, at fixed magnetic field orientation, the magnetic moment direction will be stabilized by induced magnetostrictive anisotropy along average (over thermal fluctuations) direction. At field output without it rotation, the direction of magnetic moment is stabilized by the above magnetostriction, induced in the magnetic field direction. At the input of a field of opposite direction, the magnetization reversal in such a system occurs abruptly under the influence of thermal hops over barrier due to above induced anisotropy. In other words, latter process is exactly similar to the magnetization reversal process at room temperature along easy direction.

 There is a question if the value of spontaneous magnetostriction in $CoFeB$ alloy is sufficient to explain the above effect quantitatively. For above effect to realize it is necessary that the effective field of magnetostrictive anisotropy (which appears in the granules released from matrix clamping) reach the value not less then $100\div120$ Oe, which follows from Fig.\ref{fig2}. Naturally, latter should not be a field of the anisotropy of the forced magnetostriction induced by the applied magnetic field. Rather, it should be the anisotropy field of spontaneous magnetostriction, reorientable by the applied magnetic field. The estimates of anisotropy field, related to the spontaneous magnetostriction, performed in Ref.~\onlinecite{r20} do not contradict the above value $100\div120$ Oe. Unfortunately we do not have a data about spontaneous magnetostrictive anisotropy in the alloy $Co_{0.25}Fe_{0.66}B_{0.09}$, although the alloy $Co_{0.21}Fe_{0.64}B_{0.15}$\cite{r23} has quite large magnetostrictive aspect ratio $\lambda_{s}=4.6\cdot10^{-5}$ at its magnetization up to saturation.

The temperature dependence of spontaneous magnetostrictive contribution to the anisotropy, which is always parallel to magnetization direction, is determined by two factors. First one is (temperature dependent) ``degree of release'' of granules from matrix clamping. Second one is (also temperature dependent) a probability of thermal reorientation of magnetic moment, similar to that in the case of conventional SW particles anisotropy.

\begin{figure}
\begin{center}
\includegraphics*[width=8cm]{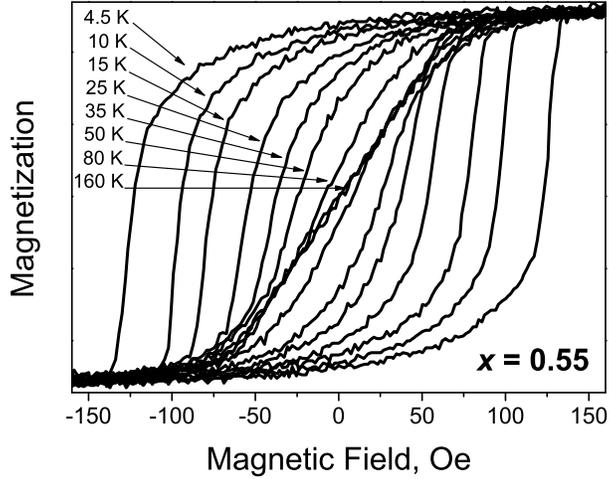}
\end{center}
\caption{\label{fig5} The set of magnetization curves for the sample with $x=0.55$ for measurements along hard (in the film plane) direction.}
\end{figure}

Let us discuss now the modification of magnetization curves for the measurements along hard (in a film plane) direction at temperature lowering. Fig. \ref{fig5} reports the family of magnetization curves (hysteresis loops) for the sample with $x=0.55$. It is seen that for high temperatures the magnetization curves have anhysteretic character with linear dependence in the scale of anisotropy field. For $T<100$ K the hysteresis appears and remanent magnetization grows. The evolution of the above magnetization curves should be considered with respect to the fact, that for the samples with $x=0.55$ and $x=0.60$ the manifestations of intergranular interaction have been observed in Ref.~\onlinecite{r18}. The latter can be well described in a mean field approximation (MFA) by addition of effective intergranular interaction field $H_{eff}^{MFA}=\lambda\cdot M$ ($\lambda$ is an interaction parameter, $M$ is an ensemble magnetization) to the external field $H$. If we consider the interaction, the anisotropy (stabilizing the magnetic moment direction in a small magnetic field) generates the coercivity. This coercivity is determined by stability limits of MFA equations\cite{r18}. The anisotropy, related to spontaneous magnetostriction (if a possibility to rotate appears for this magnetostriction) can stabilize the magnetization in a direction, which earlier was hard. For latter direction, the shape of magnetization curves is also determined by the stability limits of a state with given magnetic moment direction for $S-$shaped MFA curve $M(H)$. The shape of the curves from Fig. \ref{fig5} does not contradict to above picture.  It is seen that even at $T=4.5$ K the curve $M(H)$ is not absolutely hard. Rather, it has a smooth deflection (characteristic for MFA solutions) for magnetic moment direction along magnetic field (up to $H_{c}$) and sharp cusp after magnetization reversal.

\section{\label{sec:Verification}Verification of above model by FMR data}

\begin{figure}
\begin{center}
\includegraphics*[width=8cm]{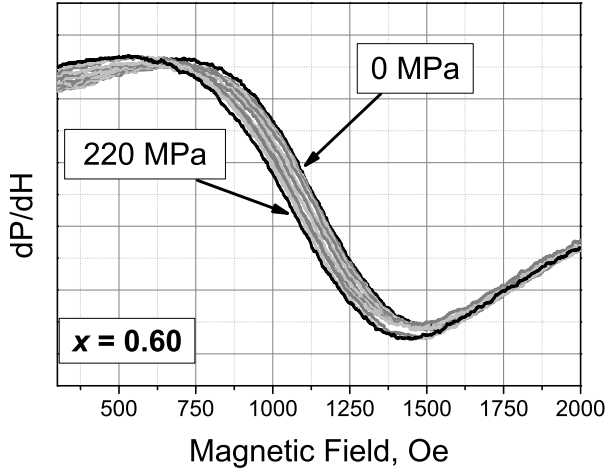}
\end{center}
\caption{\label{fig6}The set of FMR curves (the derivative of absorption curve) for the sample with $x=0.60$ at $T=295$ K with gradual enhancement of substrate pressure. The pressure was applied along easy direction and the magnetic field -- along hard (in the film plane) direction.}
\end{figure}

Seeking for additional proofs of the presence of above magnetoelastic anisotropy, we measure the ferromagnetic resonance at $T=295$ K magnetizing the sample with $x=0.60$ along hard (in a film plane) direction. In this experiment, the uniaxial compression has been applied to the sample along easy direction. The pressure was exerted not on the film itself but on a glassy substrate, on which the film has been sputtered. This means that the pressure data could not be immediately rendered into the coefficient of uniaxial compression of a film. But the experiment (see Fig. \ref{fig6}) shows the qualitative possibility of magnetoelastic anisotropy generation in a film. At the pressure enhancement the position of FMR line shifts towards smaller fields signifying the decay of initial uniaxial anisotropy (in a plane), directed perpendicular to the magnetic field or growth of anisotropy parallel to a magnetic field direction. After removal of pressure the FMR line returns to its initial position. It is seen from Fig. \ref{fig6} that under pressure 220 MPa exerted on the substrate, the FMR line shifts downwards by 100 Oe.

The FMR in granulated film had been studied many times both experimentally and theoretically. For instance, the FMR manifestations of granules and entire sample shape anisotropy for different factors of filling of a sample by granules have been considered in the papers\cite{r24,r25}. For granules concentrations, exceeding magnetic percolation threshold, the regions of inhomogeneous magnetization can appear in a film. In this case the FMR spectrum complicates\cite{r24}. For the concentration below percolation threshold, the FMR spectrum can be described by the expressions of Kittel formula type with temperature-modified magnetization. The latter quantity should be determined self-consistently by Langevin function of an argument containing resonant field. To the best of our knowledge, the temperature modification of granules anisotropy fields (do not related to their shape anisotropy) under thermally activated rotations of their magnetization has not been considered in a context of granulated films FMR neither in Refs.~\onlinecite{r24,r25} nor elsewhere in the literature. In our case the granules are amorphous and have almost spherical shape, while their planar anisotropy is correlated technologically. Thus we may suppose that there are two components of granules anisotropy, unrelated to anisotropy of their shape, namely $H_{a\theta}$ and $H_{a\varphi}$ - the axial and intraplanar components of a film as a whole. Omitting here the unnecessary details, we suppose that the above components can be considered as some effective values, averaged over certain (characteristic for FMR) ``measuring time'' in a resonant field and a given temperature. This implies that FMR can be described by formulas of Kittel type with temperature-modified magnetization $\bar{M}$ and granules anisotropy fields $\overline{H}_{a\theta}$ and $\overline{H}_{a\varphi}$. The $\overline{M}$ value is determined from the equation $\overline{M}=f\cdot M_{gS}\cdot L(V_{g}\cdot M_{gS}\cdot H_{ir}/kT)$. Here $f$ is a factor of bulk sample filling by granules, $V_{g}$ is granule volume, $H_{ir}$ is internal (with respect to demagnetizing energy) field in a film under resonant FMR field, $M_{gS}$ is a granule saturation magnetization and $L(x)$ is Langevin function. Values $\overline{H}_{a\theta}$ and $\overline{H}_{a\varphi}$ should be proportional to $H_{a\theta}$ and $H_{a\varphi}$ respectively. In general case they are temperature dependent.  In the present paper we consider them as experimentally adjustable parameters. For the measurements along easy and hard directions in a film plane the equation, relating $\overline{M}$, anisotropy fields and resonant external field $H_{res}$ on the microwave frequency $\omega$ reads

\begin{equation} \label{EQ3} 
\left(\frac{\omega}{\gamma }\right)^{2}=(H_{res}\pm2\overline{H}_{a\varphi})\cdot(H_{res}\pm\overline{H}_{a\varphi}+4\pi M_{eff}), \end{equation}

where $\gamma$ is gyromagnetic ratio,  $4\pi M_{eff}=4\pi\overline{M}(H_{ir},T)+\overline{H}_{a\theta}$. The signs ``+'' and ``-`` before terms with $\overline{H}_{a\varphi}$ correspond to the FMR measurements under magnetization along easy and hard directions in the film plane respectively. The variation of a resonant field under pressure (see Fig. \ref{fig6}) occurs from $H_{res}=1080$ Oe to $H_{res}=980$ Oe. For our estimations we use  $4\pi M_{eff}\approx9700$ Gs (this is $4\pi M$ value extracted from magnetization curves in a normal to film direction) or $8630\pm50$ Gs (value extracted from FMR measurements in the plane and along normal to a film at $T=295$ K, see below). Also, we use $(\omega/\gamma)\approx3200\div3100$ Oe, which is the value extracted from experimental $\omega$ in FMR measurements in the plane and along normal to a film. Having above quantities, we obtain anisotropy field $\overline{H}_{a\varphi}$ without pressure to be $\overline{H}_{a\varphi}(P=0)=62\div40$ Oe, while at the substrate pressure 220 MPa $\overline{H}_{a\varphi}(P=220\text{ MPa})=10\div-12$ Oe. Hence the pressure, exerted on a film surface in the above configuration reduces the magnetic anisotropy several times or even alters it sign. This confirms the magnetoelastic nature of intraplanar anisotropy of the films considered and can be regarded as an argument in favor of our model of the low-temperature anomaly in the above intraplanar anisotropy.

One more examination of our idea of magnetoelastic origin of intraplanar anisotropy of the films under consideration has been performed in the film with $x\rightarrow1$, corresponding to almost continuous $FeCoB$ film. This film with around 500 nm thickness, sputtered on glassy substrate, also demonstrates intraplanar 180-degree anisotropy for the field, lying in the film plane. We have prudently separated it from a substrate, thus obtaining a free $FeCoB$ film. For this film, the angular dependence of FMR line (at magnetic field rotation in the plane of a film) does not show the 180-degree anisotropy. This means that the above anisotropy is due to a uniaxial intraplanar strain, which appears in the process of film growth, i.e. it has magnetoelastic nature.

The aforementioned experiments prove the importance of magnetoelastic interactions in the properties of the films $(Co_{0.25}Fe_{0.66}B_{0.09})_{x}-(SiO_{2})_{1-x}$. However, they do not give direct evidence that observed low-temperature anomaly in a magnetic anisotropy of these films is related to the spontaneous granules magnetostriction. It follows from the papers \cite{r19,r20,r21} that the anisotropy (which is always directed along magnetization vector) arising from spontaneous magnetostriction, should be manifested in FMR.  The equations, obtained in Ref. ~\onlinecite{r20} for the expected (due to above anisotropy) FMR line shifts for different magnetic field (relatively to film) orientations cannot be applied directly to nanogranulated films FMR. But these equations permit to make an important conclusion. Namely, in any case the additional (due to spontaneous magnetostriction) contribution to anisotropy makes arbitrary magnetization direction to be ``easy'', i.e. it shifts the FMR line towards lower fields as compared to its position without spontaneous magnetostriction. If in our case at $T<100$ K the spontaneous magnetostriction is present, then the above shift should take place. To check that, we measure FMR in the film with $x=0.60$ in X-range of microwave frequencies and at $T=295$ and 77 K. Namely, we have measured the angular dependences of FMR lines under rotation of a magnetic field in a film plane ($\varphi$ - dependences at $\theta=\pi/2$) and out of it ($\theta$ - dependences at fixed $\varphi$). Here $\varphi$ is the angle (in a film plane) between easy magnetization and magnetic field directions and $\theta$ is the angle between field direction and normal to the film.  FMR $\theta$ - dependences have been measured on the spectrometer Bruker E 500 CW.

\begin{figure}
\begin{center}
\includegraphics*[width=8cm]{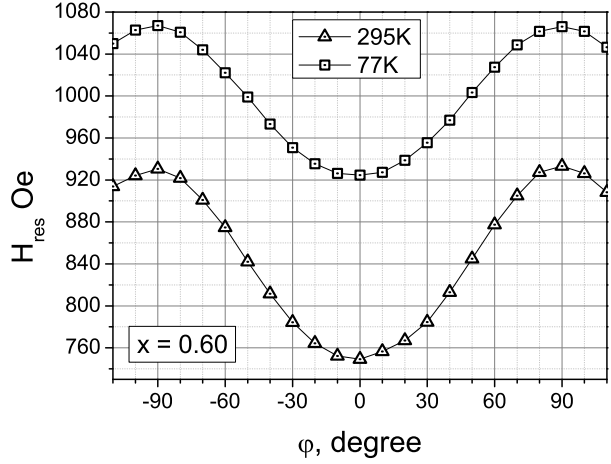}
\end{center}
\caption{\label{fig7} The angular dependences of FMR line maximum position in the film with $x=0.60$ at $T=295$ and 77 K. The experimental error corresponds to symbol size and equals to $\pm7$ Oe.}
\end{figure}

Fig. \ref{fig7} reports $\varphi$ - dependences of FMR line maximum position at $\theta=\pi/2$ in the film with $x=0.60$ for $T=295$ and 77 K. The measurements for both temperatures have been performed for the same resonant frequency 9050 MHz. The error in line maxima determination corresponds to symbol size in Fig. \ref{fig7} and equals to $\pm7$ Oe. It is seen that FMR line at $T=77$ K shifts substantially to lower fields for all angles $\varphi$ as compared to its position at $T=295$ K. The span of angular dependence is a little larger, corresponding to growth of realized intragranular anisotropy field $\overline{H}_{a\varphi}$.

At temperature variation, the shift of $\varphi$ - dependence at $\theta=\pi/2$ can be related to the dependence $4\pi M_{eff}(T)$. As it was mentioned above, the saturation magnetization of a film, extracted from magnetostatic measurements, does not alter with temperature in the region $4.5<T<295$ K. But the temperature dependence $4\pi M_{eff}(T)$ can be due to the dependence $\overline{H}_{a\theta}(T)$.

FMR $\theta$ - dependences have been measured on the microwave frequency 9169 MHz. The maxima of lines for $\theta=0$ were found to be at $H_{res}=11817\pm20$ and $13462\pm20$ Oe at $T=295$ and 77 K respectively. In the approach, similar to that for Eq. \eqref{EQ3}, the equation for $H_{res}(\theta=0)$ has the form:

\begin{equation} \label{EQ4}
\left(\frac{\omega}{\gamma }\right)^{2}=(H_{res}(\theta =0)-4\pi M_{eff})^{2}-(\overline{H}_{a\varphi})^{2}, \end{equation}
 
 It is obvious, that parameters $\overline{M}(H_{ir},T)$ and $\overline{H}_{a\varphi}(H_{ir},T)$ at fixed temperatures can be slightly different from those in Eq. \eqref{EQ3} since they should be taken for the internal fields in a sample corresponding to external resonant field. But the latter fields differ by order of magnitude for $\theta=\pi/2$ and arbitrary $\varphi$ and for $\theta=0$. However, it follows from magnetostatic measurements (see Figs. \ref{fig1}-\ref{fig4}) that for magnetization in a film plane ($\theta=\pi/2$) the film magnetization saturates beginning from $H=300\div400$ Oe so that it is already saturated for FMR resonant fields and arbitrary $\varphi$. This suggests that we can neglect the difference between $\overline{M}(H_{ir})$ and $\overline{H}_{a\varphi}(H_{ir})$ at the same temperature but in different resonant fields, corresponding (on one side) to $\theta=\pi/2$ and arbitrary $\varphi$ and $\theta=0$ on the other side. In this case, the joint solution of Eqs. \eqref{EQ3} for $\varphi=0$ and $\pi/2$ as well as Eq.\eqref{EQ4} with known resonant fields and measuring frequencies at $T=295$ K gives us $\gamma=1.8376\cdot10^{7}\text{ rad}\cdot \text{s}^{-1}\cdot\text{Oe}^{-1}$, which corresponds to $g$-factor 2.0872, $4\pi M_{eff}(T=295K)=8630\pm50$ Gs and $\overline{H}_{a\varphi}(T=295 \text{K})=39\pm3$ Oe. The substitution of obtained $\gamma$ into similar equations set for $T=77$ K does not permit to match the obtained resonant fields at $\theta=\pi/2$ at any $\varphi$ on one side and for $\theta=0$ on the other side. To match, we should either have $H_{res}(\theta=0)$ by 230 Oe higher then observed value or $\varphi$ dependence curve at $\theta=\pi/2$ should go 16 Oe higher. These discrepancies exceed experimental error. Hence, FMR lines at $T=77$ K (either all for $\theta=0$ and $\theta=\pi/2$ or only $\varphi$ dependence at $\theta=\pi/2$) are shifted to lower fields as compared to their expected (from Eqs. \eqref{EQ3} and \eqref{EQ4}) positions. This coincides qualitatively with supposition about the manifestation of spontaneous magnetostriction contribution into granules anisotropy at $T=77$ K. We hope that FMR measurements at lower temperatures would give more convincing evidence in favor of above supposition.

If we ignore the above discrepancy in the equations solutions at $T=295$ and 77 K, we can find $4\pi M_{eff}(T=77K)=10560\pm50$ Gs and $\overline{H}_{a\varphi}(T=77K)=46.5\pm$ 4 Oe.

\section{\label{sec:Conclusions}Conclusions}

We observe the effect of anomalous growth of coercive field for nanogranulated films ($Co_{0.25}Fe_{0.66}B_{0.09})_{x}-(SiO_{2})_{1-x}$  at low temperatures. To the best of our knowledge of literature, this effect has not been observed earlier. We also do not know any earlier explanation of above anomalous phenomenon in terms of magnetoelasticity with temperature variation of granule-matrix elastic coupling. The most important conclusions are following:

i) For granulated films $Co_{0.25}Fe_{0.66}B_{0.09})_{x}-(SiO_{2})_{1-x}$ with intraplanar anisotropy for $x$ below percolation threshold the temperature dependence of coercive field for the measurements in a film plane along direction of easy magnetization in the temperature range $100<T<300$ K is well described by Neel-Brown law. At $T<100$ Ê the sharp deviation of $H_{c}(T)$ from Neel - Brown law is observed. This deviation consists in faster temperature growth of $H_{c}(T)$ with temperature lowering. Simultaneously, the coercivity appears along hard (in the film plane) direction, which grows with temperature lowering. In this case, the difference curve of $H_{c}(T)$ dependences along hard and easy directions continues to satisfy Neel-Brown law. This means the appearance of additional (to the previously existing intraplanar one) anisotropy reorienting together with external magnetic field direction. At $T=4.5$ K the hysteretic curves have almost 100\% remanence both along hard (in the plane) and easy magnetization directions.

ii) Our measurements for the set of samples with different $x$ show that above low-temperature anomaly is decreased down to the complete vanishing in the samples with $x$ substantially exceeding the percolation (relatively to conductivity) threshold. Hence it is not a property of granule ferromagnetic material but rather is due to the granularity of the composite.

iii) We suppose that aforementioned effect is related to the appearance (with temperature lowering) of possibility of granules spontaneous magnetostriction and its reorientation at granules magnetic moments rotation by an external magnetic field. Here the easy axis of resulting anisotropy corresponds to that of the applied field regardless its direction in a film plane, both along previous easy and hard directions. Such possibility appears due to granules release from $SiO_{2}$ matrix clamping at temperature lowering, which is the consequence of different thermal expansion coefficients of glassy matrix and metallic granules. FMR data prove the substantial magnetoelastic influence on intraplanar anisotropy in the above material, which is qualitative argument in favor of above interpretation of the phenomenon.

Thus, despite the fact, that magnetoelastic effects in granular nanocomposites are much weaker then in a bulk ferromagnet \cite{r8}, here we observe the situation, when composite granularity generates anomalous manifestations of films magnetoelasticity at low temperatures. We do not know any previous description of the above effect in the literature. That is why we are unable to make any conclusions about ``general nature'' of observed effect. At the same time, we have no reasons to speak about its ``uniqueness'' related to ``subtleties'' of above samples fabrication technology. In our opinion, the broadening of range of low-temperature magnetostatic and magnetoresonant studies of granular composites with different thermal expansion coefficients of matrix and granules materials would permit to give an answer to the above questions.

\begin{acknowledgments}
The present work has been partially supported by Grant 10-07-N of the program of NAS of Ukraine ``Nanostructure systems, nanomaterials, nanotechnologies''. 
We are grateful to S.A. Omelchenko and A.A. Gorban for the help in organization of experiment on FMR registration under uniaxial pressure and to V.O. Golub for a possibility to perform measurements on  Bruker spectrometer.
\end{acknowledgments}

\bibliography{FeCoB-magnitostr}

\begin{thebibliography}{25}
\expandafter\ifx\csname natexlab\endcsname\relax\def\natexlab#1{#1}\fi
\expandafter\ifx\csname bibnamefont\endcsname\relax
  \def\bibnamefont#1{#1}\fi
\expandafter\ifx\csname bibfnamefont\endcsname\relax
  \def\bibfnamefont#1{#1}\fi
\expandafter\ifx\csname citenamefont\endcsname\relax
  \def\citenamefont#1{#1}\fi
\expandafter\ifx\csname url\endcsname\relax
  \def\url#1{\texttt{#1}}\fi
\expandafter\ifx\csname urlprefix\endcsname\relax\def\urlprefix{URL }\fi
\providecommand{\bibinfo}[2]{#2}
\providecommand{\eprint}[2][]{\url{#2}}

\bibitem[{\citenamefont{Munakata et~al.}(1999)\citenamefont{Munakata, Yagi, and
  Shimada}}]{r1}
\bibinfo{author}{\bibfnamefont{M.}~\bibnamefont{Munakata}},
  \bibinfo{author}{\bibfnamefont{M.}~\bibnamefont{Yagi}}, \bibnamefont{and}
  \bibinfo{author}{\bibfnamefont{Y.}~\bibnamefont{Shimada}},
  \bibinfo{journal}{IEEE Trans. Magn.} \textbf{\bibinfo{volume}{35}},
  \bibinfo{pages}{3430} (\bibinfo{year}{1999}).

\bibitem[{\citenamefont{Yamaguchi et~al.}(2008)\citenamefont{Yamaguchi, Kim,
  and Ikedaa}}]{r2}
\bibinfo{author}{\bibfnamefont{M.}~\bibnamefont{Yamaguchi}},
  \bibinfo{author}{\bibfnamefont{K.}~\bibnamefont{Kim}}, \bibnamefont{and}
  \bibinfo{author}{\bibfnamefont{S.}~\bibnamefont{Ikedaa}},
  \bibinfo{journal}{J. Magn. Magn. Mater.} \textbf{\bibinfo{volume}{304}},
  \bibinfo{pages}{206} (\bibinfo{year}{2008}).

\bibitem[{r3(2002)}]{r3}
\bibinfo{journal}{Trans. Magn. Soc. Japan} \textbf{\bibinfo{volume}{2}},
  \bibinfo{pages}{388} (\bibinfo{year}{2002}).

\bibitem[{\citenamefont{Deng et~al.}(2007)\citenamefont{Deng, Feng, J.Jiang,
  and He}}]{r4}
\bibinfo{author}{\bibfnamefont{L.}~\bibnamefont{Deng}},
  \bibinfo{author}{\bibfnamefont{Z.}~\bibnamefont{Feng}},
  \bibinfo{author}{\bibnamefont{J.Jiang}}, \bibnamefont{and}
  \bibinfo{author}{\bibfnamefont{H.}~\bibnamefont{He}}, \bibinfo{journal}{J.
  Magn. Magn. Mater.} \textbf{\bibinfo{volume}{309}}, \bibinfo{pages}{285}
  (\bibinfo{year}{2007}).

\bibitem[{\citenamefont{Kalinin et~al.}(2001)\citenamefont{Kalinin, Sitnikov,
  Stognei, Zolotukhin, and Neretin}}]{r5}
\bibinfo{author}{\bibfnamefont{Y.~E.} \bibnamefont{Kalinin}},
  \bibinfo{author}{\bibfnamefont{A.~V.} \bibnamefont{Sitnikov}},
  \bibinfo{author}{\bibfnamefont{O.~V.} \bibnamefont{Stognei}},
  \bibinfo{author}{\bibfnamefont{I.~V.} \bibnamefont{Zolotukhin}},
  \bibnamefont{and} \bibinfo{author}{\bibfnamefont{P.~V.}
  \bibnamefont{Neretin}}, \bibinfo{journal}{Materials Science and Engineering:
  A} \textbf{\bibinfo{volume}{304}}, \bibinfo{pages}{941}
  (\bibinfo{year}{2001}).

\bibitem[{\citenamefont{Wang et~al.}(2005)\citenamefont{Wang, Nordman, Qian,
  Daughton, and Myers}}]{r6}
\bibinfo{author}{\bibfnamefont{D.}~\bibnamefont{Wang}},
  \bibinfo{author}{\bibfnamefont{C.}~\bibnamefont{Nordman}},
  \bibinfo{author}{\bibfnamefont{Z.}~\bibnamefont{Qian}},
  \bibinfo{author}{\bibfnamefont{J.~M.} \bibnamefont{Daughton}},
  \bibnamefont{and} \bibinfo{author}{\bibfnamefont{J.}~\bibnamefont{Myers}},
  \bibinfo{journal}{J. Appl. Phys.} \textbf{\bibinfo{volume}{97}},
  \bibinfo{pages}{10C906} (\bibinfo{year}{2005}).

\bibitem[{\citenamefont{Johnsson et~al.}(2003)\citenamefont{Johnsson, Aoqui,
  Grishin, and Munakata}}]{r7}
\bibinfo{author}{\bibfnamefont{P.}~\bibnamefont{Johnsson}},
  \bibinfo{author}{\bibfnamefont{S.~I.} \bibnamefont{Aoqui}},
  \bibinfo{author}{\bibfnamefont{A.~M.} \bibnamefont{Grishin}},
  \bibnamefont{and} \bibinfo{author}{\bibfnamefont{M.}~\bibnamefont{Munakata}},
  \bibinfo{journal}{J. Appl. Phys.} \textbf{\bibinfo{volume}{93}},
  \bibinfo{pages}{8101} (\bibinfo{year}{2003}).

\bibitem[{\citenamefont{Cooke et~al.}(2002)\citenamefont{Cooke, Hatton, Wang,
  Szumiata, Zuberek, Watts, Gehring, and Rainforth}}]{r8}
\bibinfo{author}{\bibfnamefont{M.~D.} \bibnamefont{Cooke}},
  \bibinfo{author}{\bibfnamefont{H.~J.} \bibnamefont{Hatton}},
  \bibinfo{author}{\bibfnamefont{L.~C.} \bibnamefont{Wang}},
  \bibinfo{author}{\bibfnamefont{T.}~\bibnamefont{Szumiata}},
  \bibinfo{author}{\bibfnamefont{R.}~\bibnamefont{Zuberek}},
  \bibinfo{author}{\bibfnamefont{R.}~\bibnamefont{Watts}},
  \bibinfo{author}{\bibfnamefont{G.~A.} \bibnamefont{Gehring}},
  \bibnamefont{and} \bibinfo{author}{\bibfnamefont{W.~M.}
  \bibnamefont{Rainforth}}, \bibinfo{journal}{Phys. Rev. B}
  \textbf{\bibinfo{volume}{65}}, \bibinfo{pages}{174418}
  (\bibinfo{year}{2002}).

\bibitem[{\citenamefont{Stoner and Wohlfarth}(1948)}]{r9}
\bibinfo{author}{\bibfnamefont{E.~C.} \bibnamefont{Stoner}} \bibnamefont{and}
  \bibinfo{author}{\bibfnamefont{E.~P.} \bibnamefont{Wohlfarth}},
  \bibinfo{journal}{Philos. Trans. R. Soc. London, Ser. A}
  \textbf{\bibinfo{volume}{240}}, \bibinfo{pages}{599} (\bibinfo{year}{1948}).

\bibitem[{\citenamefont{N\`{e}el}(1949)}]{r10}
\bibinfo{author}{\bibfnamefont{L.}~\bibnamefont{N\`{e}el}},
  \bibinfo{journal}{Ann. Geophys.} \textbf{\bibinfo{volume}{5}},
  \bibinfo{pages}{99} (\bibinfo{year}{1949}).

\bibitem[{\citenamefont{Timopheev and Ryabchenko}(2008)}]{r11}
\bibinfo{author}{\bibfnamefont{A.~A.} \bibnamefont{Timopheev}}
  \bibnamefont{and} \bibinfo{author}{\bibfnamefont{S.~M.}
  \bibnamefont{Ryabchenko}}, \bibinfo{journal}{Ukr. J. Phys.}
  \textbf{\bibinfo{volume}{53}}, \bibinfo{pages}{261} (\bibinfo{year}{2008}),
  \urlprefix\url{arXiv:0803.1632v1 [cond-mat.mes-hall]}.

\bibitem[{\citenamefont{Timopheev
  et~al.}(2008{\natexlab{a}})\citenamefont{Timopheev, Kalita, and
  Ryabchenko}}]{r12}
\bibinfo{author}{\bibfnamefont{A.~A.} \bibnamefont{Timopheev}},
  \bibinfo{author}{\bibfnamefont{V.~M.} \bibnamefont{Kalita}},
  \bibnamefont{and} \bibinfo{author}{\bibfnamefont{S.~M.}
  \bibnamefont{Ryabchenko}}, \bibinfo{journal}{Low Temp. Phys.}
  \textbf{\bibinfo{volume}{34}}, \bibinfo{pages}{446}
  (\bibinfo{year}{2008}{\natexlab{a}}).

\bibitem[{\citenamefont{He and Chen}(2007)}]{r13}
\bibinfo{author}{\bibfnamefont{L.}~\bibnamefont{He}} \bibnamefont{and}
  \bibinfo{author}{\bibfnamefont{C.}~\bibnamefont{Chen}},
  \bibinfo{journal}{Phys. Rev. B} \textbf{\bibinfo{volume}{75}},
  \bibinfo{pages}{184424} (\bibinfo{year}{2007}).

\bibitem[{\citenamefont{Bedanta et~al.}(2005)\citenamefont{Bedanta, Petracic,
  Kentzinger, Kleemann, Rücker, Paul, Bruckel, Cardoso, and Freitas}}]{r14}
\bibinfo{author}{\bibfnamefont{S.}~\bibnamefont{Bedanta}},
  \bibinfo{author}{\bibfnamefont{O.}~\bibnamefont{Petracic}},
  \bibinfo{author}{\bibfnamefont{E.}~\bibnamefont{Kentzinger}},
  \bibinfo{author}{\bibfnamefont{W.}~\bibnamefont{Kleemann}},
  \bibinfo{author}{\bibfnamefont{U.}~\bibnamefont{Rücker}},
  \bibinfo{author}{\bibfnamefont{A.}~\bibnamefont{Paul}},
  \bibinfo{author}{\bibfnamefont{T.}~\bibnamefont{Bruckel}},
  \bibinfo{author}{\bibfnamefont{S.}~\bibnamefont{Cardoso}}, \bibnamefont{and}
  \bibinfo{author}{\bibfnamefont{P.~P.} \bibnamefont{Freitas}},
  \bibinfo{journal}{Phys. Rev. B} \textbf{\bibinfo{volume}{72}},
  \bibinfo{pages}{024419} (\bibinfo{year}{2005}).

\bibitem[{\citenamefont{Kleemann et~al.}(2001)\citenamefont{Kleemann, Petracic,
  Binek, Kakazei, Pogorelov, Sousa, Cardoso, and Freitas}}]{r15}
\bibinfo{author}{\bibfnamefont{W.}~\bibnamefont{Kleemann}},
  \bibinfo{author}{\bibfnamefont{O.}~\bibnamefont{Petracic}},
  \bibinfo{author}{\bibfnamefont{C.}~\bibnamefont{Binek}},
  \bibinfo{author}{\bibfnamefont{G.~N.} \bibnamefont{Kakazei}},
  \bibinfo{author}{\bibfnamefont{Y.~G.} \bibnamefont{Pogorelov}},
  \bibinfo{author}{\bibfnamefont{J.~B.} \bibnamefont{Sousa}},
  \bibinfo{author}{\bibfnamefont{S.}~\bibnamefont{Cardoso}}, \bibnamefont{and}
  \bibinfo{author}{\bibfnamefont{P.~P.} \bibnamefont{Freitas}},
  \bibinfo{journal}{Phys. Rev. B} \textbf{\bibinfo{volume}{63}},
  \bibinfo{pages}{134423} (\bibinfo{year}{2001}).

\bibitem[{\citenamefont{Kohmoto et~al.}(2004)\citenamefont{Kohmoto, Munakata,
  Mineji, and Isagawa}}]{r16}
\bibinfo{author}{\bibfnamefont{O.}~\bibnamefont{Kohmoto}},
  \bibinfo{author}{\bibfnamefont{M.}~\bibnamefont{Munakata}},
  \bibinfo{author}{\bibfnamefont{N.}~\bibnamefont{Mineji}}, \bibnamefont{and}
  \bibinfo{author}{\bibfnamefont{Y.}~\bibnamefont{Isagawa}},
  \bibinfo{journal}{Materials Science and Engineering: A}
  \textbf{\bibinfo{volume}{375}}, \bibinfo{pages}{1069} (\bibinfo{year}{2004}).

\bibitem[{\citenamefont{Omelchenko and Kulikov}(1978)}]{r17}
\bibinfo{author}{\bibfnamefont{S.~A.} \bibnamefont{Omelchenko}}
  \bibnamefont{and} \bibinfo{author}{\bibfnamefont{S.~A.}
  \bibnamefont{Kulikov}}, \bibinfo{journal}{Pribory i technika eksperimenta}
  \textbf{\bibinfo{volume}{6}}, \bibinfo{pages}{163} (\bibinfo{year}{1978}).

\bibitem[{\citenamefont{Timopheev
  et~al.}(2008{\natexlab{b}})\citenamefont{Timopheev, Ryabchenko, Kalita,
  Lozenko, Trotsenko, Stephanovich, Grishin, and Munakata}}]{r18}
\bibinfo{author}{\bibfnamefont{A.~A.} \bibnamefont{Timopheev}},
  \bibinfo{author}{\bibfnamefont{S.~M.} \bibnamefont{Ryabchenko}},
  \bibinfo{author}{\bibfnamefont{V.~M.} \bibnamefont{Kalita}},
  \bibinfo{author}{\bibfnamefont{A.~F.} \bibnamefont{Lozenko}},
  \bibinfo{author}{\bibfnamefont{P.~A.} \bibnamefont{Trotsenko}},
  \bibinfo{author}{\bibfnamefont{V.~A.} \bibnamefont{Stephanovich}},
  \bibinfo{author}{\bibfnamefont{A.~M.} \bibnamefont{Grishin}},
  \bibnamefont{and} \bibinfo{author}{\bibfnamefont{M.}~\bibnamefont{Munakata}},
  \bibinfo{journal}{Submitted to JAP}  (\bibinfo{year}{2008}{\natexlab{b}}),
  \urlprefix\url{arXiv:0805.2463v2 [cond-mat.mes-hall]}.

\bibitem[{\citenamefont{Borovik-Romanov and Rudashevskii}(1965)}]{r19}
\bibinfo{author}{\bibfnamefont{A.~S.} \bibnamefont{Borovik-Romanov}}
  \bibnamefont{and} \bibinfo{author}{\bibfnamefont{E.~G.}
  \bibnamefont{Rudashevskii}}, \bibinfo{journal}{JETP}
  \textbf{\bibinfo{volume}{20}}, \bibinfo{pages}{1407} (\bibinfo{year}{1965}).

\bibitem[{\citenamefont{Turov and Shavrov}(1965)}]{r20}
\bibinfo{author}{\bibfnamefont{E.~A.} \bibnamefont{Turov}} \bibnamefont{and}
  \bibinfo{author}{\bibfnamefont{V.~G.} \bibnamefont{Shavrov}},
  \bibinfo{journal}{Sov. Phys. Solid State} \textbf{\bibinfo{volume}{7}},
  \bibinfo{pages}{166} (\bibinfo{year}{1965}).

\bibitem[{\citenamefont{Borovik-Romanov
  et~al.}(1984)\citenamefont{Borovik-Romanov, Rudashevskii, Turov, and
  Shavrov}}]{r21}
\bibinfo{author}{\bibfnamefont{A.~S.} \bibnamefont{Borovik-Romanov}},
  \bibinfo{author}{\bibfnamefont{E.~G.} \bibnamefont{Rudashevskii}},
  \bibinfo{author}{\bibfnamefont{E.~A.} \bibnamefont{Turov}}, \bibnamefont{and}
  \bibinfo{author}{\bibfnamefont{V.~G.} \bibnamefont{Shavrov}},
  \bibinfo{journal}{Sov. Phys. Uspekhi} \textbf{\bibinfo{volume}{27}},
  \bibinfo{pages}{642} (\bibinfo{year}{1984}).

\bibitem[{\citenamefont{Gann and Zhukov}(1982)}]{r22}
\bibinfo{author}{\bibfnamefont{V.~V.} \bibnamefont{Gann}} \bibnamefont{and}
  \bibinfo{author}{\bibfnamefont{A.~I.} \bibnamefont{Zhukov}},
  \bibinfo{journal}{Sov. Phys. Solid State} \textbf{\bibinfo{volume}{24}},
  \bibinfo{pages}{1584} (\bibinfo{year}{1982}).

\bibitem[{\citenamefont{Kraus and Svec}(2003)}]{r23}
\bibinfo{author}{\bibfnamefont{L.}~\bibnamefont{Kraus}} \bibnamefont{and}
  \bibinfo{author}{\bibfnamefont{P.}~\bibnamefont{Svec}}, \bibinfo{journal}{J.
  Appl. Phys.} \textbf{\bibinfo{volume}{93}}, \bibinfo{pages}{7220}
  (\bibinfo{year}{2003}).

\bibitem[{\citenamefont{Netzelmann}(1990)}]{r24}
\bibinfo{author}{\bibfnamefont{U.}~\bibnamefont{Netzelmann}},
  \bibinfo{journal}{J. Appl. Phys.} \textbf{\bibinfo{volume}{68}},
  \bibinfo{pages}{1800} (\bibinfo{year}{1990}).

\bibitem[{\citenamefont{Kakazei et~al.}(1999)\citenamefont{Kakazei, Kravets,
  Lesnik, de~Azevedo, Pogorelov, and Sousle}}]{r25}
\bibinfo{author}{\bibfnamefont{G.~N.} \bibnamefont{Kakazei}},
  \bibinfo{author}{\bibfnamefont{A.~F.} \bibnamefont{Kravets}},
  \bibinfo{author}{\bibfnamefont{N.~A.} \bibnamefont{Lesnik}},
  \bibinfo{author}{\bibfnamefont{M.~M.~P.} \bibnamefont{de~Azevedo}},
  \bibinfo{author}{\bibfnamefont{Y.~G.} \bibnamefont{Pogorelov}},
  \bibnamefont{and} \bibinfo{author}{\bibfnamefont{J.~B.}
  \bibnamefont{Sousle}}, \bibinfo{journal}{J. Appl. Phys.}
  \textbf{\bibinfo{volume}{85}}, \bibinfo{pages}{5654} (\bibinfo{year}{1999}).

\end{thebibliography}

\end{document}